\documentclass{sig-alternate-nocopyright}

\usepackage{graphicx}
\usepackage[utf8]{inputenc}
\usepackage[numbers]{natbib}
\usepackage{listings}
\usepackage[T1]{fontenc}
\usepackage[english]{babel}
\usepackage{xspace}
\usepackage{color}
\usepackage{hyperref}
\usepackage{algpseudocode}
\usepackage{algorithm}

\newcommand{\redo}{{\small REDO}\xspace}
\newcommand{\undo}{{\small UNDO}\xspace}

\begin{document}

\title{A novel recovery mechanism enabling \\ fine-granularity locking and fast, REDO-only recovery}

\numberofauthors{2} 

\author{
 \alignauthor Caetano Sauer\\
 \affaddr{University of Kaiserslautern} \\
 \affaddr{Germany} \\
 \email{csauer@cs.uni-kl.de} 
 \alignauthor Theo H\"arder\\
 \affaddr{University of Kaiserslautern} \\
 \affaddr{Germany} \\
 \email{haerder@cs.uni-kl.de}
}

\date{\today}

\maketitle

\begin{abstract}
We present a series of novel techniques and algorithms for transaction commit, logging, recovery, and propagation control.
In combination, they provide a recovery component that maintains the persistent state of the database (both log and data pages) always in a committed state.
Recovery from system and media failures only requires only \redo operations, which can happen concurrently with the processing of new transactions.
The mechanism supports fine-granularity locking, partial rollbacks, and snapshot isolation for reader transactions.
Our design does not assume a specific hardware configuration such as non-volatile RAM or flash---it is designed for traditional disk environments.
Nevertheless, it can exploit modern I/O devices for higher transaction throughput and reduced recovery time with a high degree of flexibility.
%So far, our mechanism is tailored to centralized DBMSs.
\end{abstract}

\section{Introduction}
The recovery component of a database system is responsible for ensuring atomicity and durability of transactions (the ``A'' and ``D'' of ACID).
It achieves these properties by carefully logging modifications to data cached in volatile memory (i.e., pages in the buffer pool) and controlling their propagation to the persistent database.
System designs for a recovery component can be classified according to a well-known taxonomy of policies for logging, propagation control, and checkpointing \cite{bib:HaerderReuter83}, leading to a variety of possible combinations.

Despite the large spectrum of design choices, the ARIES mechanism \cite{bib:Mohan92ARIES} has become a \emph{de-facto} standard in traditional disk-based architectures.
ARIES implements a \emph{no-force}/\emph{steal} policy, meaning that (i) it does not require pages to be flushed back to disk at commit time and (ii) pages can be flushed at arbitrary times, even if they contain uncommitted updates.
These policies are considered crucial to achieve high transaction throughput and fast recovery times in case of failures.
However, the consequence of this design choice is that transaction recovery requires both \undo and \redo logging, following the write-ahead logging (WAL) principle \cite{bib:JimGrayNotes,bib:RecoverySystemR}.
While WAL has become ubiquitous in practice, especially following ARIES, many older proposals exist which either reduce logging or eliminate it altogether.

In contrast to ARIES, a \emph{force}/\emph{no-steal} policy eliminates the need for logging and provides instantaneous recovery at the cost of substantially increased commit latencies.
Given the unacceptable performance penalty, traditional designs which offer an alternative to WAL employ a combination of techniques to provide simplified (or absent) logging and faster recovery \cite{bib:BayerDBCache,bib:HaerderTOSP}.
However, these approaches have not been implemented in modern database systems for two reasons: (i) they achieve faster recovery at the expense of significantly slower transaction processing; and (ii) they require page-level locking, thus severely limiting concurrency among transactions.
The success of ARIES is largely due to the fact that it solves these two problems.

This paper presents a design for a recovery component supporting fine-granularity locking with faster and easier recovery than ARIES.
By carefully controlling which updates are allowed to reach persistent storage, we eliminate the need for \undo recovery after a system or media failure.
Because updates of loser transactions never reach either the database or the persistent log, only \redo operations are required for recovery.
This is enabled by two main techniques:
\begin{itemize}
  \item A commit protocol that employs private \undo logs in volatile memory for transaction rollback. Log records of commited transactions are copied to the persistent log in a single atomic step.
  \item A propagation control technique that maintains persistent database pages always in a committed state by undoing updates of uncommitted transactions on a single page.
    %Our approach is based on the idea of single-page recovery \cite{bib:SinglePageRecovery}.
\end{itemize}

The novel design enables a simpler implementation of logging, eliminating the need for compensation log records and \emph{undo-next} pointers.
Checkpoints are also simplified as a consequence of the absence of \undo recovery.
Based on the design of \emph{instant restart} \cite{bib:InstantRestart}, new transactions can be admitted shortly after restart following a system failure, i.e., before the database is fully recovered.
Furthermore, it enables simpler implementation of other techniques and utilities which rely on the recovery component, such as snapshot isolation, temporal databases, and online backups.

Most importantly, our design works for system architectures based on both disk/flash and non-volatile RAM.
As memory latency decreases, our design can be simply tuned to provide higher throughput and faster recovery, maintaining the same base algorithm regardless of storage media.
Currently, our design is tailored to centralized DBMSs, but possible extensions for NUMA or cluster architectures are conceivable.

The remainder of this paper is organized as follows.
Section \ref{sec:CommitProtocol} presents the commit protocol that maintains the persistent log in a committed state.
Conversely, Section \ref{sec:Propagation} presents a propagation control scheme for maintaining the persistent database in a committed state.
Section \ref{sec:VLM} discusses techniques to implement a volatile log manager, which is the basic underlying component supporting our design.
Section \ref{sec:Recovery} discusses checkpoints and recovery from system failures, 
Section \ref{sec:Applications} analyzes the impact of our design on applications other than recovery, whereas Section \ref{sec:Modern} discusses the applicability of our solution to modern hardware architectures, such as multi-socket and multi-core CPUs, as well as non-volatile memory.
Finally, Section~\ref{sec:Conclusion} concludes this paper.

\section{Atomic Commit Protocol}\label{sec:CommitProtocol}
The major goal of our design is to eliminate the need for \undo recovery after a failure.
The first step in that direction is to maintain the persistent log in a committed state.
To that end, we propose a commit protocol that holds log records of active transactions in volatile memory, copying them to the centralized log at commit time in a single atomic step.
The first important distinction to ARIES is that there are two types of logs: one \emph{persistent} log and multiple per-transaction \emph{volatile} logs.
In this scenario, the sole purpose of the persistent log is to ensure durability, whereas volatile logs provide atomicity.

\subsection{Per-transaction volatile logs}
The WAL protocol requires that every update to a page (which happens in-place in its buffer pool image) is preceded by the generation of a log record that allows the change to be undone or redone.
ARIES employs the principle of physiological logging \cite{bib:GrayReuter}, whereas older implementations rely on before- and after images of pages \cite{bib:Bernstein87}.
What virtually all implementations have in common is that both \undo and \redo information are collected in the same centralized, persistent log.
In our approach, log records are initially placed in a volatile log which is private to the transaction that generated them.
Recent proposals for logging and recovery on non-volatile RAM rely on the same technique \cite{bib:StorageMgmtNVRAMEra,bib:Johnson14LoggingNVRAM}, albeit for different purposes.
We shall discuss them later in Section~\ref{sec:Modern}.

Transaction rollback simply requires a backward scan over the private log of the aborting transaction.
Since in this case the log records are placed adjacent to each other in memory, there is no need for \emph{undo-next} pointers, which are used in ARIES to find the previous log record of the same transaction.
Undoing a logged update also does not require a compensation log record (CLR), which is used in ARIES to ensure the correct behavior of \undo actions under fine-granularity locking and repeated crashes.
Using volatile logs for \undo means that a rollback can only happen as a consequence of a transaction abort (e.g., because of deadlocks) and not of crash recovery.
This implies that the anomalies which CLRs were designed to avoid cannot occur.

\subsection{Log propagation}
When a transaction commits, its log records must be synchronously copied into the persistent log.
To guarantee the atomicity property, a transaction is only considered committed (i.e., a ``winner'') if all its log records---including the commit log record---reach persistent storage.
The ARIES design relies on a centralized log buffer where log records are inserted for every update.
To commit a transaction, a commit log record is inserted and the log buffer is simply flushed to persistent storage.
In our design, there is no distinction between the \emph{insert} and \emph{flush} operations in the centralized log---the only operation supported is an atomic \emph{copy} of all log records from a transaction's private log.
For implementation purposes, the private logs themselves can be viewed as log buffers for the copy operation\footnote{This assumes that log records always contain both \undo and \redo information.}.

The copy operation must be synchronized so that log records of the same transaction are placed contiguously in the persistent log.
This means that only one transaction can commit at a time.
In ARIES, a single log flush operation ends up flushing the log records of multiple transactions, but in the end this does not make a difference for commit contention, since a log flush must finish before the next transaction can commit.
Therefore, we expect the same overall contention and I/O penalty during commit processing as in ARIES.
In both cases, the group commit technique \cite{bib:GrayReuter} is strongly recommended to allow multiple transactions to commit in a single log flush and thus enable higher transaction throughput.

% TODO move this to later discussion of Johnsons work, read Aether paper agian
%A disadvantage of ARIES is that the log insert operation must be synchronized among threads to protect the physical integrity of log records.
%Approaches such as Aether \cite{bib:Johnson10Aether} minimize the insert contention by synchronizing only the phases of space reservation in the log buffer and the subsequent release, allowing the actual filling (i.e., copy) to occur in parallel among multiple threads.
%Despite being efficient, the scheme is quite complex to implement, mainly because of log space reservation issues.
%In our atomic commit design, the contention is minimal by design, because log records are inserted into private transaction logs.
% TODO this discussion depends on whether a log buffer should be used or not -- add this discussion somewhere

\subsection{LSN and pointer assignment}\label{sec:LSNassign}
Given that log records are initially produced locally for each transaction, there is no centralized mechanism to generate unique LSNs at the time log records are generated.
Instead, LSNs are generated at commit time.
When log records are copied to the centralized log, their byte offset in the persistent log file is assigned as the LSN value.
This is the most common implementation option for LSNs, since a byte offset gives direct access to log records and thus can be used as a pointer.
Assigning LSNs when a log record is generated is not a viable solution, because the position that the log record will occupy in the log file (or whether it will end up in the file at all) is not known initially.
To implement this solution, the private log assigns initial LSN values as byte offsets within the private transaction log.
At commit time, the transaction is assigned an offset in the persistent log.
Before log records are flushed, this global offset is added to the local LSNs previously assigned, yielding the final LSN value.

The concept of a PageLSN, which is used to indicate the version of a page on persistent storage, must be redefined for atomic commits.
In ARIES, every update to a page also sets the PageLSN value to the LSN of the log record generated for that update.
When a page is flushed, the PageLSN value indicates a point in the log where all previous updates are already reflected in the page image.
By comparing this value with the LSN of log records during recovery, only the required \undo and \redo operations are performed.
In our design, however, the PageLSN may only be updated at commit time.
This is required because only committed updates reach persistent storage and, thus, the PageLSN must reflect the state of a page after the updating transaction committed.
Furthermore, as a consequence, updates of aborted transactions do not modify the PageLSN of a page.
%Given these crucial distinctions, we shall use the term \emph{version} instead of PageLSN in the reminder of this paper.

The procedure for assignment of LSNs to log records and pages is given in Algorithm \ref{alg:lsn}.
It takes as arguments a transaction identifier $t$ and the amount $s$ of log space required.
Note that the exact value of $s$ is known at commit time, since all log records were already generated.
The procedure updates the LSN value of each log record based on the assigned offset in the persistent log.
It also updates the per-page log chain and the PageLSN value of the modified pages.

\begin{algorithm}
    \begin{algorithmic}[1]
    \Procedure{AssignLSNs}{t,s}
      \State $offset\gets reserveSpace(s)$ \label{alg:lsn:reserve}
      \ForAll {$rec$ in $privateLog(t)$}
          \State {$rec.lsn\gets rec.lsn + offset$} \label{alg:lsn:setlsn}.
          \If{$rec.pageID \neq null$} \label{alg:lsn:testpage}
              \State $p\gets rec.pageID$
%                \State $lockVersion(p, commit\_duration)$ \label{alg:lsn:lock}
              \State $latchPage(p)$
              \State $rec.prev\gets getPageLSN(p)$ \label{alg:lsn:prev}
              \State $setPageLSN(p, rec.lsn)$ \label{alg:lsn:setpagelsn}
              %\State $setUnpropagated(p)$ \label{alg:lsn:setunprop}
              \State $unlatchPage(p)$
          \EndIf
      \EndFor
    \EndProcedure
    \end{algorithmic}
    \caption{Procedure for LSN assignment\label{alg:lsn}}
\end{algorithm}

The first step of the algorithm is to reserve space in the persistent log, which returns an offset in the log file where the $s$ bytes will be copied.
Based on this offset, the actual LSN value of each log record is set (line \ref{alg:lsn:setlsn}).
For log records that entail page updates (i.e., which contain a valid page ID), the per-page log chain must be updated based on the current PageLSN value of the page (line \ref{alg:lsn:prev}.
After that, the PageLSN value must be updated, which requires access to the page.
Thus, its physical integrity must be protected by acquiring a latch, which also guarantees that the page will not be flushed in an intermediary state in between.
Note that the page must be fetched if it has been evicted from the buffer pool, but this situation can be avoided by a proper page replacement strategy.

In order to avoid latching all updated pages to increase their PageLSN at commit time, which essentially doubles the contention on exclusive page latches, some optimizations are conceivable.
%For example, a \emph{pending LSN} field can be added to each buffer control block.
%During commit, the $setPageLSN$ function simply sets the pending LSN to the new PageLSN of the page in the corresponding buffer control block.
%When the page is fixed by a further update operation, the buffer manager checks the pending LSN field and updates the PageLSN field accordingly before returning from the fix method.
%Note that this still incurs contention on the buffer hash table and on buffer control blocks, but the page latch itself is left untouched.
One option is to implement the $setPageLSN$ operation in an indirect manner, using a separate table mapping page IDs to their PageLSN values.
To allow for high concurrency, a latch-free hash table could be used \cite{bib:LatchFreeHashTable}, eliminating the need to acquire expensive page latches.
The feasibility of this table approach comes from the observation that PageLSNs are only accessed during system-related activities, and never directly by user transactions
\footnote{This includes B-tree operations that must check whether a page was updated since it was last read, as done, for instance, in ARIES/KVL\cite{bib:MohanARIESKVL}. We refer back to this problem in Section \ref{sec:VLM}}.
To that end, the PageLSN value of an image in the buffer pool could be simply ignored, and every read or update should instead be directed to the separate table.
When a page is selected for propagation it is then important to set the correct PageLSN value on the disk image.
Another option is to decouple PageLSN accesses from normal page accesses by using separate latches, thus eliminating the conflict between user transaction threads and system threads that perform commits, page cleaning, and recovery.
%Line \ref{alg:lsn:prev} sets the \emph{prev} pointer of log records, which points to the log record of the previous update on the same page, building a so-called \emph{per-page log chain}.
%This enables the implementation of single-page recovery and many other useful features \cite{bib:SinglePageRecovery}.
%In our design, it is required for \redo recovery after a system failure as well as for propagation control using single-page \redo.
%We shall return to these points later in Section \ref{sec:Propagation}, which describes the relationship between the commit protocol and propagation control.

%Once LSNs and pointers are properly set, the log records are copied to the persistent log and the private log can be recycled.
%These steps concludes the atomic commit protocol.

Once the procedure for LSN assignment is completed, the private log records are copied into the reserved space and a successful commit message is returned.

\subsection{System transactions}
The commit protocol presented so far assumes that all log records reflect only updates to user data.
However, the maintenance of internal structures for data representation (e.g., B-tree structure or space management) also requires transactional consistency.
A classical example is a page split in a B-tree.
One of the virtues of ARIES' record locking mechanism is that a transaction $t_1$ can split a B-tree node and immediately after allow another transaction $t_2$ to access and modify data in the new page created by the split, even if $t_1$ must be rolled back later.
This is accomplished with \emph{top-level actions} \cite{bib:Mohan92ARIES}.

Top-level actions as defined by ARIES are not directly supported in the atomic commit protocol, because they require the mechanism of \emph{undo-next} pointers and CLRs.
Instead of trying to accommodate top-level actions in our design, a much cleaner solution is to implement \emph{system transactions} \cite{bib:GraefeBTreeLogging}, which are a special kind of lightweight transaction that only modifies the physical representation of data.
System transactions can be implemented in the atomic commit protocol in two ways.
One possibility is to insert their log records directly into the centralized log, simply ignoring them if a commit log record is not found during recovery after a system failure.
Just like user transactions, \undo is not required because uncommitted changes cannot reach persistent storage.
Recall that this is a fundamental assumption of the atomic commit protocol.
Section \ref{sec:Propagation} discusses ways to ensure this property.

A second implementation alternative is to use a special private log exclusive to system transactions.
When a user transaction goes through the commit protocol, it can simply copy the committed log records in the private log of system transactions together with its own log records.
To identify which log records are committed, it looks for the latest commit log record and removes all preceding log records.

\section{Propagation Control}\label{sec:Propagation}
A propagation control module is responsible for bringing transaction updates, which initially happen only in the buffer pool, into the persistent database.
The atomic commit protocol discussed so far only yields a correct implementation of a recovery component if flushed pages contain only updates of committed transactions.
Therefore, the propagation control module must guarantee that the pages it flushes contain only updates of committed transactions.
Traditionally, this guarantee is provided by means of a \emph{no-steal} policy \cite{bib:HaerderReuter83}.
``Stealing'' a page means to flush its contents back to the persistent database while it still contains updates of at least one active transaction.
Hence, if stealing is not permitted, uncommitted updates will not be propagated.

The major drawback of a no-steal policy---and possibly the reason why it is not often implemented in practice---is that a page flush operation must wait for any active transactions to finish and quiesce all updates on that page until the flush is completed.
The second main contribution of our design---next to the atomic commit protocol---is a propagation control mechanism which achieves no-steal guarantees without quiescing updates, by means of \undo logging and rollback.
Before introducing the technique in detail, we must first discuss propagation control in traditional \emph{steal} implementations such as ARIES.

\subsection{Page flushing in ARIES}
The key characteristic of a steal policy is that pages can be selected for flushing regardless of transaction activity on them.
However, this does not mean that a page can be flushed at any arbitrary point in time---the physical consistency must still be protected by acquiring a latch \cite{bib:Mohan92ARIES}.
This means that the ``stealing'' only happens at the level of logical consistency, i.e., there is no need to acquire database locks.

In order to minimize the overhead of flushing a page, a copy is taken into a separate in-transit buffer and the latch is immediately released.
The write operation is then invoked from the in-transit buffer.
This procedure provides propagation control in accordance with the general guideline of not holding latches during I/O operations \cite{bib:Mohan92ARIES}.

The consequence of a steal policy for recovery is that \undo logging is required to roll back, during restart after a system failure, the updates of transactions which did not commit.
By applying such rollback operations at the time a page is flushed, we can achieve the effect of a no-steal policy without the associated overhead.

%If system transactions are supported, the no-steal policy allows the atomic commit protocol to be implemented with very few changes to the original ARIES algorithm \cite{bib:Mohan92ARIES}.
%The necessary changes are summarized below:
%\begin{itemize}
%  \item Replace top-level actions with system transactions
%  \item Implement private volatile logs for each (reusable) transaction object.
%  \item Implement a \emph{copy} method in the log manager, taking as argument a reference to a transaction's private log.
%  \item Replace all invocations of the log manager's \emph{insert} method with a simple append to the transaction's private log.
%  \item For transaction commit, implement Algorithm \ref{alg:lsn} and replace the call to \emph{flush} with a call to \emph{copy}.
%\end{itemize}

\subsection{Single-page rollback}
We propose a propagation control module based on a technique of \emph{single-page rollback}, which we abbreviate as SPR (together with \emph{single-page replay}, to be introduced in Section \ref{sec:Recovery}).
The inspiration for SPR comes from the idea of single-page recovery \cite{bib:SinglePageRecovery}, which is a mechanism that allows individual pages to be recovered if they become corrupt, e.g., due to flash wear-out or controller errors.
Our approach builds upon the principle of single-page recovery to support the \undo of updates on an individual page.

We follow the ARIES approach of copying pages to be flushed into an in-transit buffer.
However, instead of flushing the page contents \emph{as is}, we first restore the page back to its latest committed state, by undoing all updates of active transactions.
If the page is already in a committed state, the rollback is not required and the flush operation is performed directly, as in ARIES.
In essence, our propagation control strategy follows the principle of \emph{flushing only committed pages}.

Note that the PageLSN value is not updated during single-page rollback.
This is because only committed changes increase PageLSN values, and in SPR only uncommitted changes are rolled back.
By applying SPR to a page image, we essentially restore it to the state determined by its PageLSN.
Recall from Section \ref{sec:CommitProtocol} that a PageLSN value is valid only for page images in committed state.

We present the procedure for page flushing in Algorithm \ref{alg:flush}.
The rollback is applied by using an iterator which delivers all uncommitted updates to a given page earlier than a given LSN.
The implementation of this iterator (i.e., the function $getUndoChain$ in the algorithm) is discussed in Section \ref{sec:VLM}. 

\begin{algorithm}
    \begin{algorithmic}[1]
    \Procedure{FlushPage}{pid}
        \State{$latchPage(pid)$}
        \State{$copy \gets copyPage(pid)$}
        \State{$unlatchPage(pid)$}
        \State{$plsn \gets getPageLSN(copy)$}
        \State{$iter \gets getUndoChain(pid, plsn)$}
        \While {$iter.hasNext()$}
            \State {$logrec \gets iter.next()$}
            \State{$undo(rec, copy)$}
        \EndWhile
        \State{$write(pid, copy)$}
    \EndProcedure
    \end{algorithmic}
    \caption{Procedure for page flushing\label{alg:flush}}
\end{algorithm}

\subsection{Page life cycle}
\begin{figure*}
  \centering
  \includegraphics[width=0.8\linewidth]{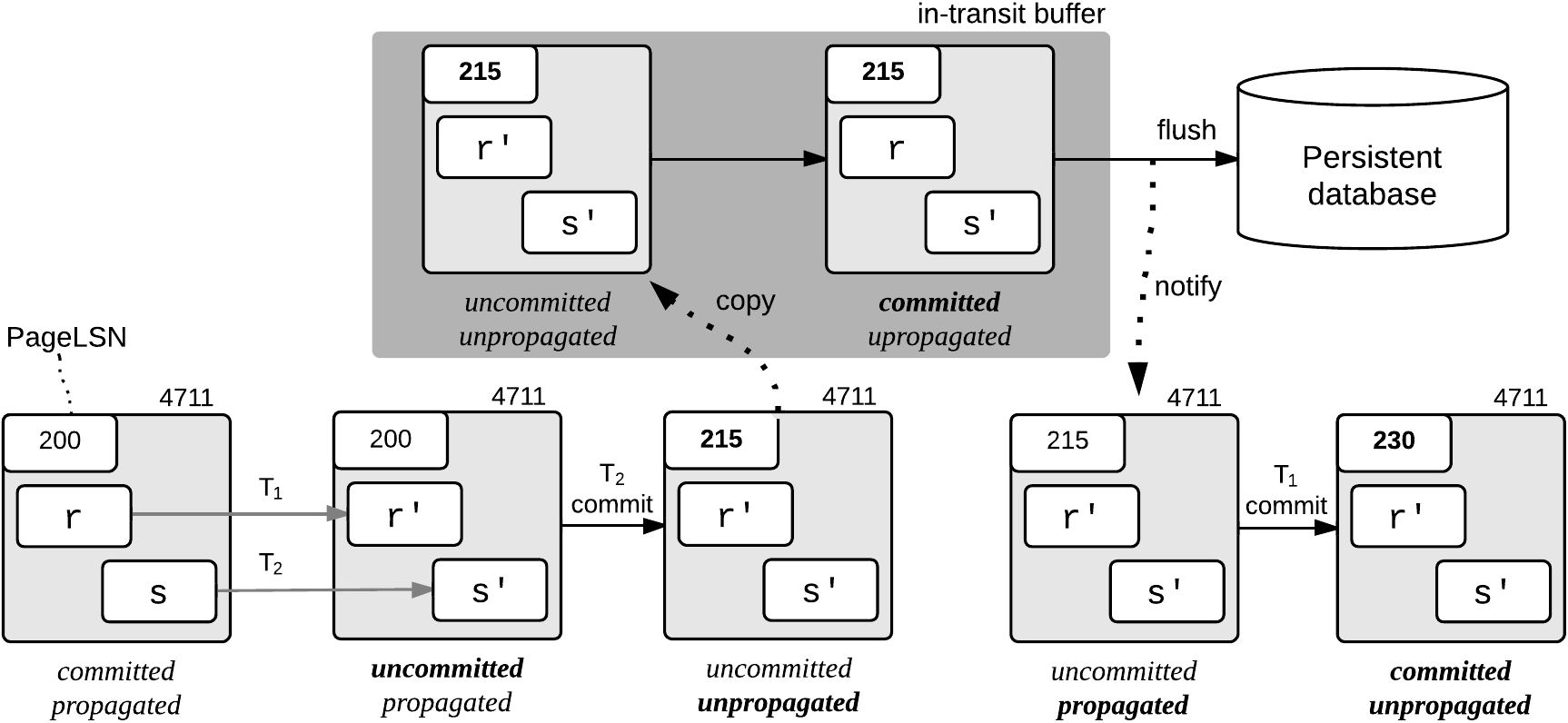}
  \caption{Life cycle of a page going through updates, commits, and propagation}\label{fig:PageLifeCycle}
\end{figure*}

In order to summarize the steps involved in the atomic commit protocol and in propagation control using SPR, we present a sequence of operations on a page which demonstrates all possible states in its life cycle.
It also introduces definitions that replace the ARIES concept of dirty/clean.
The flowchart in Figure \ref{fig:PageLifeCycle} starts with a page with ID 4711 which was just fetched into the buffer pool with PageLSN 200.
At this stage, the page is in a \emph{committed} state, since it may not contain updates of uncommitted transactions due to the SPR protocol.
Its contents in the buffer pool are also the same as in the persistent database (e.g., on disk), which means it is in a \emph{propagated} state.

Next, two transactions, $T_1$ and $T_2$, update records $r$ and $s$, respectively, bringing them to new states $r'$ and $s'$.
At this stage, the page is in an \emph{uncommitted} state, since the updating transactions have not committed yet.
However, it is still in the \emph{propagated} state, even though the actual page image differs from persistent storage.
This is because propagation control involves single-page rollback.
If page 4711 is selected for propagation at this point, the records $r$ and $s$ would be restored, leaving the page in the same state as on persistent storage and rendering the propagation unnecessary.
Another way of defining the \emph{propagated} state is by relying on the PageLSN value.
Note that after the updates it has the same value as when it was fetched---if it would differ, the page would be in an \emph{unpropagated} state.

Next, transaction $T_2$ commits, updating the PageLSN to 215 and leaving the page in the \emph{unpropagated} state.
However, it still remains \emph{uncommitted} because $T_1$ is still active.
At this stage, the page is selected for propagation.
A copy is taken and, because it is in the \emph{uncommitted} state, SPR is invoked, which rolls back the update of $T_1$.
Once SPR completes, the in-transit copy is on a \emph{committed} state, making it eligible for flushing.
After the flush completes, the buffer manager is notified and the page in the buffer pool is marked as \emph{propagated}, but only if its PageLSN did not change since the in-transit copy was taken, which is the case in this example.
This is equivalent to how pages are marked clean in ARIES.
Finally, $T_1$ commits and the page is again in a \emph{committed} state with PageLSN 230.
If it is selected for propagation at this point, no rollback is necessary.

\section{Volatile Log Manager Design}\label{sec:VLM}
From an architectural perspective, we propose splitting the traditional log manager into two subcomponents: a \emph{persistent} log manager and a \emph{volatile} log manager (VLM).
The former is involved in the commit protocol introduced in Section \ref{sec:CommitProtocol}, and its main responsibility is the provision of an atomic $copy$ method for flushing log records of a private transaction log into the persistent log.
We omit the implementation details of this subcomponent because of its relative simplicity in comparison with the latter subcomponent, the volatile log manager.

In order to perform the actual rollback, there must be an efficient way of retrieving the volatile log records pertaining to a given page ID.
The most important design goal in this process is that such log records can be retrieved with minimal interference in transaction-processing activity, mainly volatile log record insertion, commit processing, and transaction abort.
We present two possible design choices for a volatile log manager, followed by a discussion of their pros and cons.

As a common design basis, we assume that the volatile log manager owns a contiguous address space which it divides into \emph{extents} of fixed size.
When a transaction starts, it must \emph{allocate} an extent to insert its log records.
In case of overflow, additional extents are allocated and linked together by means of \emph{next-extent} pointers.
At some point in time, depending on the end state of the owning transaction and recycling policies, extents are freed by the VLM itself.
This means that the volatile log manager also embodies the functionality of a simple memory manager with garbage collection.

\subsection{Using log chains}
Finding log records that pertain to a given page ID is also a requirement in the single-page recovery design \cite{bib:SinglePageRecovery}, which is based on traditional ARIES-style logging (i.e., centralized and persistent \undo and \redo logging).
The technique employed is a per-page log chain, in which log records ``point'' to the previous update on the same page using an LSN value.
Despite not being directly applicable to volatile log records, the same principle can be applied.
Therefore, our first design proposal introduces the notion of per-page chains of volatile logs.

The first observation on volatile log chains is that the chaining mechanism must be substantially different from the one employed during recovery, because log records reside in volatile private logs.
The first difference is that LSN values---which are used as pointers for recovery---cannot be used, because their values are assigned only at commit time.
Therefore, we introduce the notion of \emph{volatile log record address} (VLA), which represents an in-memory address of a log record in a volatile log.
Given the basic memory management design given earlier, VLAs can be assigned as byte offsets within the VLM extent address space\footnote{Note that unlike LSNs, VLAs do not need to preserve chronological order, because they are used only for addressing and not in comparisons.}.
By introducing a VLA field into a log record's header, newly-inserted volatile log records can point to the previous volatile log record whose update affected the same page.
The head of the chain can be stored in a VLA field inside a page's buffer control block.
In the recovery case, the current PageLSN is used as head of the chain, but in the volatile case there is no need to store the field inside the page contents, since it does not have to be persisted (i.e., it is not required for crash recovery).
Whenever a page is updated and a volatile log record is generated, the head VLA is copied into the pointer field and the head updated with the new VLA.

Using the volatile per-page log chain, a copy of a page selected for flushing can be restored to its latest committed state, but this process is not straight-forward, as the following questions need to be answered:

\begin{enumerate}
  \item At which point in the chain must the rollback begin, i.e., what is the first update to be undone?
  \item Single-page rollback must employ \emph{selective \undo}, i.e., undoing only uncommitted updates, but how can we determine (without blocking transaction progress) whether or not a log record qualifies for \undo when navigating the per-page chain?
  \item At which point in the chain must the rollback stop, i.e., how to determine whether the latest committed state was reached?
  \item How to determine when volatile log records can be deleted by garbage collection in the VLM?
\end{enumerate}

The first problem is easily solved because single-page rollback starts from a copy of a page in the buffer pool.
When the copy is taken, the VLA of its latest log record can be extracted from the field in the buffer control block.
Race conditions do not occur because the page latch held during the copy also protects the contents of the page's control block.
Even if further updates occur after the copy is taken, the head of the chain is preserved, pointing to the correct volatile log record.
This assumes, for now, that the chain of pointers never changes during runtime, except for adding new log records to the head and removing log records from the \emph{tail}, which we discuss in the context of Question 3.

To determine which updates committed and which not, the set of active transactions must be known.
Instead of asking the transaction manager for a ``snapshot'', which would probably require some form of expensive latching, this could be achieved by making use of extent metadata in the VLM.
If each extent contains a \emph{committed} flag, there is no need to communicate with the transaction manager.
However, two issues must be addressed.
First, the access to the flag (and the extent metadata in general) must be synchronized so that either all updates of a transaction are undone or none of them.
Note that undoing updates of a transaction that already committed (i.e., false positives) is not harmful, as it simply leaves the page at an older (but consistent) state.
Therefore, transactions can freely commit as long as they latch and update their extents when doing so.
The second issue arrives when a transaction's log spans multiple extents.
To provide atomicity of selective \undo, such transactions must have all extents latched for the entire duration of a single-page rollback.
This is also required when the metadata is updated by a committing transaction.
Furthermore, such latches must be acquired in a fixed order to avoid deadlocks.
Since handling multiple-extent transactions is expensive, the extent size should be big enough to accommodate the vast majority of transactions in the workload.

Transactions may freely insert log records into an extent while it is latched by the rollback procedure.
However, undoing log records on uncommitted extents may slow down transactions entering the commit phase, because they must latch the extent in order to update the committed flag.
This contention can be significantly reduced by simply waiting a short amount of time after the copy of the page is taken.
If the workload consists mostly of short transactions, there is a high likelihood that all transactions in the chain will have finished (with either commit or abort) by the time single-page rollback starts.
In this case, navigating the chain will have absolutely no effect on transaction throughput, because latches of committed/aborted extents are only acquired by single-page rollback.

To answer the third question, we must determine where the rollback must stop, i.e., the chain's \emph{tail}.
One possible solution is to ``trim'' the chain at each pass during single-page rollback, updating the tail at each page flush.
When a page is allocated, its first log record contains a $nil$ pointer, which indicates the chain tail.
When the page is flushed, its chain is navigated until the $nil$ pointer is found, but the rollback mechanism keeps track of the last log record which was undone.
When the tail is reached, the pointer in that log record can be set to $nil$, since all log records in between are already committed and thus must not be undone by future page flushes.

The fourth problem is closely related to the third.
The solution presented for updating the chain tail has the major disadvantage that log records must be kept in volatile memory for a relatively long time---until all pages touched by a transaction are cleaned.
Furthermore, garbage collection would be cumbersome because it would only be able to free an extent once all its log records have been invalidated by the tail update mechanism, which would be expensive to keep track of.
Therefore, we propose a more proactive garbage collection mechanism that removes committed log records from the middle of the chain to which they belong.
By doing that, extents could be freed as soon as a transaction commits, and the per-page chains would be kept very short---possibly empty if the waiting technique is employed.

One disadvantage of this garbage collection mechanism is that the update of a chain must obviously synchronize with the navigation of that chain.
This can be achieved by properly latching extents.
When removing a log record $B$ from a partial chain $A\rightarrow B\rightarrow C$, the extents of $A$ and $B$ must be latched.
In order to preserve atomicity of selective \undo, extents must be freed atomically in a per-transaction basis, i.e., all extents of a transaction must be latched.
If extents are always latched in an ordered fashion (e.g., most recent to least recent), this mechanism safely blocks single-page rollback from seeing an inconsistent pointer chain.

The main advantage of the VLM design based on log chains is that it scales well, since latches are acquired only at the extent level.
By employing the waiting technique and avoiding multiple-extent transactions, the contention is minimized, and no noticeable impact on transaction throughput is expected.
However, the need to maintain the per-page chains, possibly updating them for efficient garbage collection, requires a cumbersome latching protocol, which makes the approach hard to understand and to implement correctly, especially in order to test all possible conflicts.
Therefore, we propose an alternative design which is much simpler to implement but may be worse in terms of scalability.

\subsection{Using volatile log sequence numbers}
If per-page log chains are not available, the only way to find all log records pertaining to a certain page is to search the logs of all active transactions.
However, this introduces synchronization issues between active transactions inserting volatile log records and single-page rollback scanning their private logs.
Before discussing these issues below, we first introduce the mechanism used to deal with them: a \emph{volatile log sequence number}, or VLSN.
Instead of using VLAs to identify volatile log records, we make use of a single atomic counter for the whole log manager.
Every time a volatile log record is generated for an update, this counter is atomically incremented and its value is used as the VLSN.
Furthermore, a field in the buffer control block of a page contains the counter value of the last update made to that page, i.e., its $PageVLSN$.

When a page is selected for propagation, its PageVLSN is copied from the buffer control block---this will be used as an upper-bound for the search, since only log records with a lower VLSN must be inspected for rollback.
To find the volatile log records pertaining to the selected page, each private log must be scanned from the begin up to the lowest value greater than the PageVLSN.
To avoid reading a malformed log record which is currently being inserted by another thread, each private log must be protected by a latch, which means a transaction cannot make any progress while its log is being scanned by single-page rollback.
To alleviate this contention, we propose the waiting technique: if the rollback mechanism waits a short amount of time, most transactions that were running when it started will likely have committed or aborted.
To exploit this fact, we make use of a $StartVLSN$ field for each private log, which simply contains the global counter value at the time the transaction started.
Then, single-page rollback only inspects a private log if its StartVLSN is lower than the given PageVLSN.

The PageVLSN field can also be used by mechanisms that require checking if a page changed since the last time it was latched, such as the B-tree concurrency protocol of ARIES/KVL \cite{bib:MohanARIESKVL}.

The main advantage of the VLSN design is its simplicity, as there is no need to maintain pointers among volatile log records.
The only concern with this method is its scalability.
Because a global atomic counter has to be read and incremented very often, it may become a source of contention.
However, this contention is only an issue in highly parallel multi-core environments \cite{bib:Johnson14LoggingNVRAM} and, at worst, it is as high as current high-performance ARIES-based designs such as Aether \cite{bib:Johnson10Aether}, where insertion at the centralized log buffer is the main source of contention.

\subsection{Discussion}
The two designs presented offer different trade-offs in terms of simplicity and scalability.
In a traditional centralized memory architecture (i.e., single-socket, non-NUMA machines), we believe the potential contention on the global counter is not an issue, and, therefore, the VLSN design should be preferred.
However, as multi-socket CPUs and non-volatile RAM proliferate, the global counter design is likely to become a bottleneck, in which case the log-chain approach may become favorable.

\section{Checkpoints and Recovery}\label{sec:Recovery}
The recovery procedure in our design is very similar to ARIES, except that certain steps and collected information are eliminated.
During crash recovery, only the phases of log analysis and \redo pass are required.
Note that the concept of repeating history, as opposed to selective \redo, is not applicable, because every update redone during recovery is from a ``winner'' transaction.
Therefore, additional optimizations such as ``restricted repeating of history'' \cite{bib:ARIES-RRH} are unnecessary.

The absence of \undo during recovery greatly simplifies not only the algorithms for crash and media recovery, but also for checkpoints.
In ARIES and most modern implementations, checkpoints are of the \emph{fuzzy} type, meaning that they do not write any database page \cite{bib:Mohan92ARIES}.
Instead, they only collect information about which pages are dirty in the buffer pool (i.e., pages that need \redo in case of a crash), and active transactions (i.e., transactions whose updates require \undo in case of failure).
Since uncommitted changes do not reach persistent storage, there is nothing to undo during recovery, and thus checkpoints need to collect only dirty pages or, in our terminology, \emph{unpropagated} pages.

%Furthermore, based on the idea of single-page recovery \cite{bib:SinglePageRecovery}, new transactions can be admitted immediately after log analysis, performing the required \redo actions on demand, as pages are requested by running transactions.
%This idea, which we call \emph{instant recovery}, is made possible by the per-page chain on log records, a technique common in modern DBMSs and described in the context of our design in Section \ref{sec:CommitProtocol}.

The technique of \emph{instant restart} \cite{bib:InstantRestart} can also be easily applied to our design.
It allows new transactions to be admitted immediately after the log analysis phase, provided that locks of loser transactions are reacquired and that a per-page log chain is available in the recovery log.
The required \redo and \undo steps are then applied on demand, as new transactions touch data that needs recovery (i.e., in-doubt pages or conflicting locks held by loser transactions).
The main advantage of our design is that instant restart, like many other components and utilities of the database system, can be substantially simplified due the absence of uncommitted data on persistent storage.
If no \undo is required for recovery, there is no need for lock reacquisition during log analysis or for a lock-conflict detection mechanism to trigger transaction rollbacks.
Further simplifications can also be envisioned for the case of recovery from media failures.

%\section{Further Uses and Applications}\label{sec:Applications}
%Besides providing fast, simplified, \redo-only recovery, the fact that all persistent data is in a committed state enables a wide variety of features and simplifications on related database components and utilities.
%In this section, we provide a brief overview of some envisioned applications.

\section{Snapshot isolation \& time travel}\label{sec:Applications}
The SPR mechanism provides a straight-forward way of implementing snapshot isolation for reader transactions.
The idea is to make copies of pages and perform the necessary rollback/replay to restore the committed state as of begin of transaction.
This allows a long-running reader transaction to simply acquire a latch while the page is being copied, without the need for database locks which may block updating transactions.
The obvious trade-off here is to increase concurrency at the cost of increased memory usage.
%The technique is probably better suited for workloads with few long-running transactions (e.g., table scans) or many short write-intensive transactions.

With respect to logging, rollback to an arbitrary page version also requires \undo information to be kept in the persistent log, even though it is never used during recovery.
One potential advantage of the atomic commit protocol is that log volume may be reduced by storing only \redo information.
However, this would require either a conversion of log records to their \redo-only versions during commit or two private logs for each transaction---one for \undo and one for \redo, i.e., the former for aborting and the latter for committing.
Because this makes the design more cumbersome to implement, we believe that maintaining combined \undo and \redo information is the better option.

The protocol for restoring a page to a point in time given by LSN $l$ is to first copy the page and roll back all updates of active transactions, as done in page flushing for SPR.
Then, the per-page chain is followed in the persistent log until a log record $r_1$ with LSN lower than $l$ is reached.
It cannot be guaranteed that the after-image of $r_1$ is in a state of transactional consistency, because the same transaction may update the same page multiple times, in log records $r_1 \cdots r_n$.
To cope with this problem, $r_1 \cdots r_n$ must all be undone.
Note that these log records will be adjacent to each other in the per-page chain, because the commit protocol atomically inserts all log records of a transaction in the persistent log at commit time.

The log-based snapshot isolation mechanism can in principle also be implemented in ARIES, but in a more cumbersome way.
After rolling back all updates greater than $l$, including both active and finished transactions, there may still be uncommitted updates left in the page, namely those of transactions that started before $l$.
Furthermore, the updates of the same transaction are not guaranteed to be adjacent in the per-page chain.
The implication is that a \emph{selective \undo} of all active transactions at time $l$ must be applied.
In an ARIES-based recovery log, the only way to determine active transactions is to perform a log scan from $l$ to the closest checkpoint log record, either in forward or backward direction.
Once the set of active transactions at time $l$ is determined, the per-page log chain is followed and selective \undo is applied until the oldest LSN in the set is reached.

Given the complexity involved, the original ARIES mechanism is not suitable for a simple implementation of log-based snapshot isolation.
The atomic commit protocol, on the other hand, provides guarantees and facilities that make the implementation straight-forward.

The mechanism of snapshot isolation can be generalized for ``time traveling'', i.e., accessing an arbitrary old database state, enabling the support for temporal databases.
At the page access level, a \emph{fix} operation can be implemented, which takes an arbitrary LSN as argument and performs the rollback operations above.
Reader transactions in snapshot isolation (or time travelling) mode then simply pass the current LSN at the time they started to every page access.

\section{Modern Hardware}\label{sec:Modern}

\subsection{Logging to NVM}
The emergence of non-volatile memory (NVM) changes fundamental assumptions of the logging and propagation control schemes of ARIES, which was designed to cope with traditional disk-based architectures.
In this section, we investigate the applicability of our design to NVM log devices, further extending the discussion to NVM for the database pages in Section \ref{sec:DBonNVRAM}.

% TODO talk about how to make log flushes atomic on NVM (Pelley's persist_wal)

If the private transaction logs discussed in Section \ref{sec:CommitProtocol} are stored in NVM, then log records need not be copied into a centralized persistent log in order to commit a transaction.
To determine after a system failure what updates need to be undone or redone, the recovery mechanism must rely on state information of the volatile log manager (VLM) introduced in Section \ref{sec:VLM}.
Recall that private logs are stored in fixed-length memory extents managed by the VLM, where extents belonging to the same transaction are linked together.
During commit, a private transaction log can be marked as committed in a single atomic step, for example by setting a committed bit on the first extent.
The other steps of the atomic commit protocol---LSN assignment and maintenance of the per-page chain---are still required for correct \redo.
In this case, however, the LSN can be generated by a simple monotonic counter, since there is no centralized recovery log where all log records end up.
The per-page chain must then use the in-memory addresses instead of LSNs to refer to the previous update on the same page.

Recovery from a system failure follows the same protocol presented in Section \ref{sec:Recovery}.
The difference is that the log analysis phase must scan the extents of the VLM instead of a centralized recovery log.
Since log analysis is only concerned with determining in-doubt pages, i.e., unpropagated pages, the order in which the extents are analyzed does not matter, as long as only committed extents are considered.
Moreover, checkpoint information must be kept in a dedicated extent which is processed in the beginning of log analysis.
One additional change required by this mechanism is that the recycling of log extents must be guided by propagation of pages, essentially in the same way that the centralized recovery log performs recycling.
An alternative mechanism relies on a log archiving daemon that periodically moves old log records into a separate storage location, which is used for media recovery.

This recovery scheme based on private logs eliminates two of the most severe bottlenecks of traditional ARIES design.
First, it is a form of distributed logging, which as argued in recent related work \cite{bib:Johnson14LoggingNVRAM}, is essential for scalability on multi-socket (i.e., NUMA) hardware.
Second, it eliminates the I/O operation required during commit processing in order to flush log records to stable storage, thus substantially lowering transaction latency and lock retention.
Unlike approaches recently proposed for logging and recovery on modern storage devices \cite{bib:StorageMgmtNVRAMEra}, our design employs the same basic techniques for both disk-based and NVM-based architectures, with only minor adaptations to the algorithms.

\subsection{Database pages on NVM\label{sec:DBonNVRAM}}
Based on the assumption that NVM devices are large enough to replace disks in a DBMS and that they are fast enough to replace (at least partially) DRAM, current proposals for recovery and propagation algorithms introduce radical changes to the established ARIES algorithms.
The basic assumption of such proposals is that the buffer manager, at least in its traditional role, can be ``skipped'', because the actual propagation of updates can be as fast as the update themselves in DRAM.
In this section, we briefly survey some of these proposals, arguing that an in-depth study is required in order to properly evaluate new approaches.

Pelley et al. propose a mechanism of \emph{in-place updates}, in which updates are made directly on persistent storage \cite{bib:StorageMgmtNVRAMEra}.
The scheme is designed for NVM devices with very low latencies, comparable to DRAM.
In order to provide atomicity, an \undo log containing before-images of pages (or fragments of a page) is employed, which is essentially equivalent to an ARIES write-ahead log.
The scheme performs well for DRAM latencies, but becomes worse than ARIES as the latency increases.
To cope with such higher latencies, the authors propose a \emph{group commit} mechanism, in which updates of transactions are propagated in batches using a force policy, quiescing transactions while propagation is performed.
The mechanism employs a \emph{staging buffer} in which uncommitted pages are kept, which is essentially a buffer manager where pages are cached only when updated.
Again, atomicity of propagation is provided by means of an \undo log.
Despite being more efficient, cumbersome techniques are required to deal with long transactions and large staging buffers.
For this reason, we argue that eliminating WAL altogether is not an ideal solution, and small modifications to ARIES should be considered instead of complete redesigns (i.e., evolution instead of revolution).

The H-Store in-memory database employs a propagation scheme based on traditional write-ahead logging and on-line transaction-consistent snapshots \cite{bib:HStoreRecovery}, similar in principle to System R \cite{bib:RecoverySystemR}.
The difference is that such snapshots are taken using a record-oriented copy-on-write mechanism, instead of using shadow paging.
The advantage is that transaction activity is not blocked while a snapshot is being taken.
Given a transaction-consistent persistent state, logical logging can be applied to substantially decrease the overhead of WAL \cite{bib:HaerderReuter83}.
H-Store uses a very efficient form of logical logging, referred to as \emph{command logging}, in which only stored procedure identifiers are logged together with their parameters.
Further studies of the H-Store team show that this architecture also performs better than ARIES in an NVM-based architecture \cite{bib:HStoreNVM}.
However, the authors conclude that the speedup reduces significantly as workload skew increases, and, therefore, they argue that a new propagation and recovery scheme is required for NVM architectures, i.e., existing disk- or memory-oriented mechanisms are not suitable.

One disadvantage of the memory-oriented logging scheme of H-Store \cite{bib:HStoreRecovery} is that recovery takes much longer due to logical logging.
Not only that, because the propagation scheme is based on transactions instead of pages, it does not permit instant restart \cite{bib:InstantRestart}.
Because of these two reasons, the H-Store mechanism provides poor availability in the presence of failures.

Given the disadvantages of the proposals discussed here, we argue that an ARIES-based solution (write-ahead logging with page-oriented propagation) may be also suitable for NVM architectures.
Our design is ARIES-based in which it does not change its fundamental assumptions.
We believe that simple, incremental adaptations to the algorithm may substantially improve its performance, making it competitive with NVM-tailored designs.
Such possibility will be further investigated in our future work.

\subsection{Scalability of logging operations}
Recent hardware trends and scalable storage manager designs have identified logging as one of the major bottlenecks for scalability on multi-core CPUs \cite{bib:Johnson09ShoreMT}.
The Aether design \cite{bib:Johnson10Aether} proposes a series of optimizations that gradually remove scalability bottlenecks from traditional I/O-bound implementations of the ARIES mechanism.
In this section, we address the scalability concerns of the atomic commit protocol and analyze the applicability of the techniques developed in Aether.

As a strategy to minimize lock contention due to transactions waiting for a log flush to complete, the technique of \emph{early lock release} is suggested.
It allows transactions to release locks after they generate a commit log record, but before its log records are flushed to persistent storage, thus substantially reducing the duration for which locks are held.
In order to reduce the scheduling overhead induced by multiple threads waiting for I/O during log flushes, Aether introduces the idea of \emph{flush pipelining}.
It allows worker threads to continue work for other transactions once a commit happens, delegating the I/O wait to a dedicated log flush daemon thread, which essentially eliminates the need for expensive context switches during commit.
Both techniques are directly applicable to the atomic commit protocol---locks may be released before log records are moved from volatile to persistent log and the move itself may be pipelined by making use of a flush daemon.

Further optimizations introduced by Aether are based on the assumption that log records are inserted into a centralized in-memory log buffer, as suggested by the ARIES design.
This means that the log buffer may become a major source of contention in highly parallel scenarios.
In our design, the insertion of log records into a volatile log is not a source of contention, because it happens in private per-transaction logs.
The $copy$ operation, on the other hand, introduces contention to the centralized persistent log, and thus it is worthwhile to investigate the applicability of ideas from Aether to our design.

Aether proposes a \emph{consolidation array} to minimize the contention of reserving space in the log.
The idea is that multiple requests for space reservation---which implies the assignment of LSN values---can be composed into a single large request.
The same idea can be applied in our design, by allowing space reservations, as shown in Algorithm \ref{alg:lsn}, to group requests of multiple threads.

A second technique, referred to as \emph{decoupled buffer inserts}, allows the operation of filling the reserved space with the contents of a log record to happen concurrently, only making sure that the reserved areas are \emph{released} in the proper LSN order to maintain correctness of recovery.
The same general idea can be applied to our design, but a substantial difference is that the requested buffer space is on a persistent device, which essentially unifies the concepts of filling and flushing.
After reserving space, log records can be flushed from private volatile logs in any order, or even in parallel, if supported by the I/O device.
However, in order to provide durability, flush requests must be returned in strict LSN order.
Such functionality can be delegated to the log flush daemon, which can emit commit messages to clients as the high boundary of safely committed LSNs increases.

We conclude that the atomic commit protocol is able to exploit the same scalability opportunities as a classical ARIES design, as done in Aether \cite{bib:Johnson10Aether}.
At a first glance, it is hard to predict whether our design scales better than ARIES, but since there are no additional sources of contention in the log manager, we do not expect reduced scalability.
%There is of course the issue of PageLSN updates, as discussed in Section \ref{sec:LSNassign}, but in that case the concern is the contention on data pages, and not on logging operations.

\section{Conclusion and Future Work}\label{sec:Conclusion}
We propose new algorithms for recovery and propagation control.
In combination, they are able to maintain the persistent state of the database---including data pages and the log---always in a committed state, thus eliminating the need for \undo recovery.
This is achieved without any major performance penalty in comparison to ARIES.
In the future, we expect to implement the algorithms in a real system and compare their performance to an ARIES-based implementation, both in normal transaction processing and during recovery.

Other than simplifying logging and recovery, we believe our algorithms may further enable or simplify other database features and utilities.
As an example, we discussed how to implement log-based snapshot isolation and time travelling, which can be achieved in a much simpler way than in ARIES.
We also expect the algorithms to facilitate backup and log archiving for media recovery.
Furthermore, if fast NVM devices are available, the commit latency can be greatly decreased by simply using persistent memory for the private transaction logs.
% TODO maybe remove all NVM ``advantages'' of paper

\section*{Acknowledgements}
We are deeply grateful to Goetz Graefe for the regular exchange of ideas and intellectual support to our research. His ideas on instant recovery provided the insights that lead to the design proposed here.
%We also thank Ryan Johnson and Pinar T\"{o}z\"{u}n, who generously and quickly answered our many questions about Shore-MT.

\bibliographystyle{spbasic}
\bibliography{recovery}

\end{document}